\documentclass[10pt,twocolumn,english]{article}
\usepackage[T1]{fontenc}
\usepackage{latex8}
\usepackage{times}
\usepackage{array}
\usepackage{booktabs}
\usepackage{amsmath}
\usepackage{graphicx}
\usepackage{amssymb}

\newcommand{\noun}[1]{\textsc{#1}}
\providecommand{\tabularnewline}{\\}

\usepackage{latex8}
\usepackage{times}

\begin{document}

\title{Making Hand Geometry Verification System More Accurate Using Time
Series Representation with R-K Band Learning}

\author{Vit Niennattrakul\\
\and Chotirat Ann Ratanamahatana}

\affiliation{Department of Computer Engineering, Chulalongkorn University\\Phayathai
Rd., Pathumwan, Bangkok 10330 Thailand}

\email{\{g49vnn, ann\}@cp.eng.chula.ac.th}

\maketitle
\thispagestyle{empty}
\begin{abstract}
At present, applications of biometrics are rapidly increasing due
to inconveniences in using traditional passwords and physical keys.
Hand geometry, one of the most well-known biometrics, is implemented
in many verification systems with various feature extraction methods.
In recent work, a hand geometry verification system using time series
conversion techniques and Dynamic Time Warping (DTW) distance measure
with Sakoe-Chiba band has been proposed. This system demonstrates
many advantages, especially ease of implementation and small storage
space requirement using time series representation. In this paper,
we propose a novel hand geometry verification system that exploits
DTW distance measure and R-K band learning to further improve the
system performance. Finally, our evaluation reveals that our proposed
system outperforms the current system by a wide margin, in terms of
False Acceptance Rate (FAR), False Rejection Rate (FRR), and Total
Success Rate (TSR) at Equal Error Rate (EER).
\end{abstract}

\Section{Introduction}

Nowadays, biometrics is gradually used in place of traditional passwords
and physical keys simply because the passwords could be forgotten
and the physical keys could be lost or stolen. On the other hand,
many parts of human organs or characteristics such as a fingerprint,
hand geometry, an iris, voice, a face profile, and a signature, are
unique enough to be used for person identification purposes. Specifically,
Table \ref{Flo:table1} shows a comparison among various biometrics
based on seven attributes \cite{Jain05,JainRP06}, i.e., universality,
uniqueness, permanence, collectability, performance, acceptability,
and circumvention.

\begin{table*}
\caption{Various biometric comparison based on seven attributes \cite{JainRP06}}

\noindent \centering{}\begin{tabular}{|c||c|c|c|c|c|c|c|}
\hline 
\textbf{\footnotesize Biometrics} & \textbf{\footnotesize Universality} & \textbf{\footnotesize Uniqueness} & \textbf{\footnotesize Permanence} & \textbf{\footnotesize Collectability} & \textbf{\footnotesize Performance} & \textbf{\footnotesize Acceptability} & \textbf{\footnotesize Circumvention}\tabularnewline
\hline
\hline 
\textbf{\footnotesize Face} & {\footnotesize High} & {\footnotesize Low} & {\footnotesize Medium} & {\footnotesize High} & {\footnotesize Low} & {\footnotesize High} & {\footnotesize Low}\tabularnewline
\hline 
\textbf{\footnotesize Fingerprint} & {\footnotesize Medium} & {\footnotesize High} & {\footnotesize High} & {\footnotesize Medium} & {\footnotesize High} & {\footnotesize Medium} & {\footnotesize High}\tabularnewline
\hline 
\textbf{\footnotesize Hand geometry} & {\footnotesize Medium} & {\footnotesize Medium} & {\footnotesize Medium} & {\footnotesize High} & {\footnotesize Medium} & {\footnotesize Medium} & {\footnotesize Medium}\tabularnewline
\hline 
\textbf{\footnotesize Keystroke} & {\footnotesize Low} & {\footnotesize Low} & {\footnotesize Low} & {\footnotesize Medium} & {\footnotesize Low} & {\footnotesize Medium} & {\footnotesize Medium}\tabularnewline
\hline 
\textbf{\footnotesize Iris} & {\footnotesize High} & {\footnotesize High} & {\footnotesize High} & {\footnotesize Medium} & {\footnotesize High} & {\footnotesize Low} & {\footnotesize High}\tabularnewline
\hline 
\textbf{\footnotesize Signature} & {\footnotesize Low} & {\footnotesize Low} & {\footnotesize Low} & {\footnotesize High} & {\footnotesize Low} & {\footnotesize Medium} & {\footnotesize Low}\tabularnewline
\hline 
\textbf{\footnotesize Voice} & {\footnotesize Medium} & {\footnotesize Low} & {\footnotesize Low} & {\footnotesize Medium} & {\footnotesize Low} & {\footnotesize Medium} & {\footnotesize Low}\tabularnewline
\hline
\end{tabular}\label{Flo:table1}
\end{table*}

A biometric authentication system can be categorized into two main
functionalities \cite{JainRP06} \textendash{} identification and
verification. For identification, a system receives an image from
an input sensor, extracts features, and queries from a database for
the best match. If the input image is too different from the best
retrieved template, the system rejects. Otherwise, it accepts and
identifies the input's identity. Unlike identification, a verification
system requires a user identity as an additional input. Instead of
querying from all the templates in the database, the verification
system queries on only the claimed user's own templates. Matching
only on much smaller set of templates generally makes the verification
system much more accurate than the identification system.

\begin{figure}
\noindent \begin{centering}
\begin{tabular}{cc}
\includegraphics[width=3.5cm]{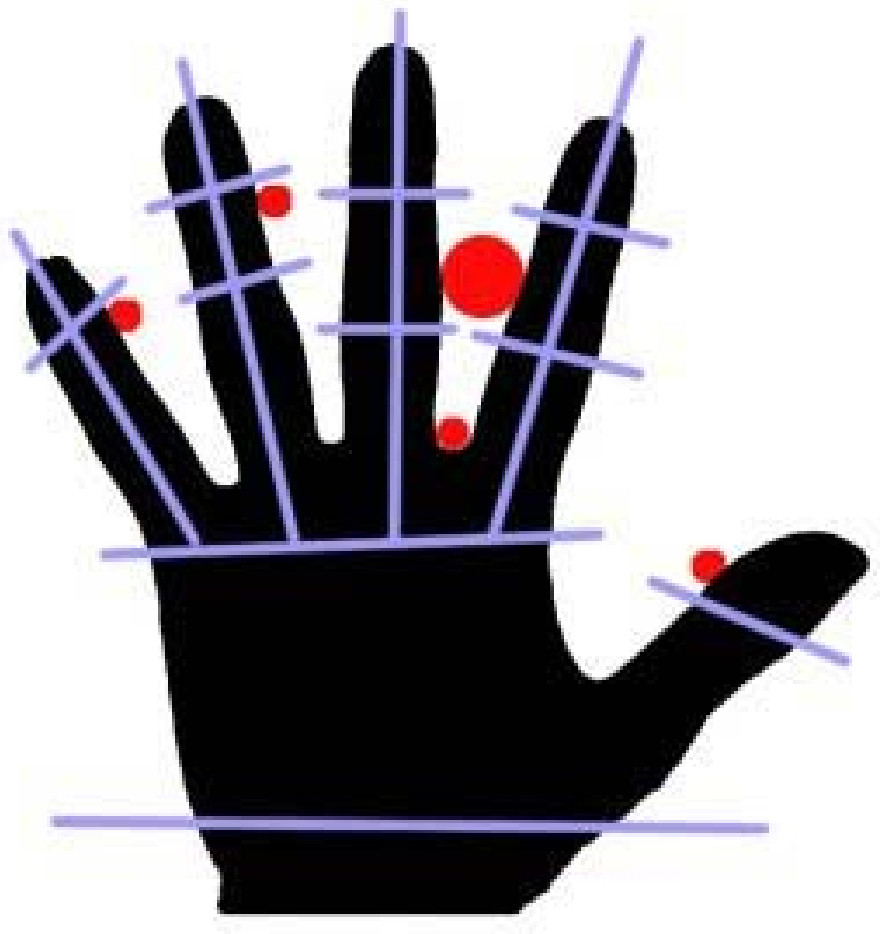} & \includegraphics[width=3.5cm]{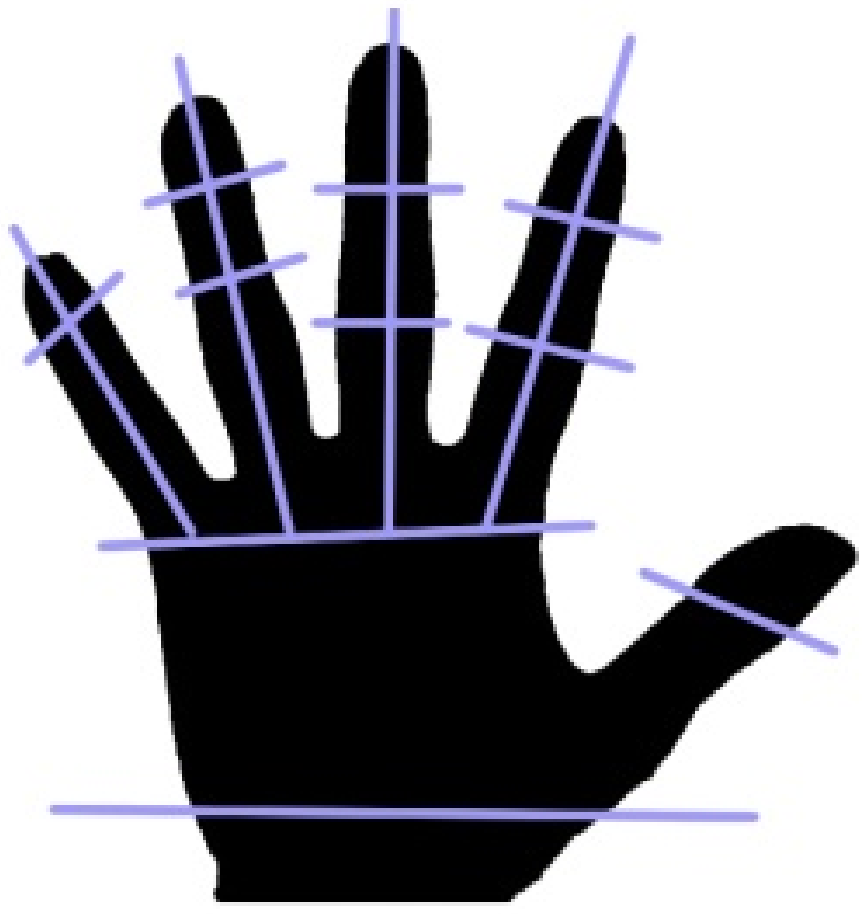}\tabularnewline
(a) & (b)\tabularnewline
\end{tabular}
\par\end{centering}

\caption{Images acquired from (a) a peg-fixed system and (b) a peg-free system.}
\label{Flo:fig1}
\end{figure}

At present, many hand geometry authentication systems are implemented
both in identification and verification tasks due to the ease of sample
collection, relatively low hardware costs, and availability of various
feature extraction algorithms. Specifically, two system environments
are designed, shown in Figure \ref{Flo:fig1}, i.e., a peg-fixed system
\cite{825671,1594879,stasiak:634725,m7450k6311n82730} and a peg-free
system \cite{4129340,1630119,Faundez-ZanuyM05,1297573,982526,KumarWSJ03,1703170,4263249,YorukKSD06,OdenEYKB01,BorekiZ05}.
The peg-fixed system uses pins attached on a plate to align hand's
position properly. On the other hand, in the peg-free system, no pin
is needed. To transform an input image to computable numerical features,
many extraction methodologies have been proposed \cite{KumarWSJ03,YorukKSD06,BulatovJKS04,JainD99,Sanchez-ReilloSG00,OdenEB03,citeulike:1633072}.
Unfortunately, these extracted features techniques are quite computationally
expensive and the system may fail to extract features in some of the
hand configurations. 

Recently, a hand geometry verification system using a time series
representation \cite{NiennattrakulWR07} has been proposed. Some of
its advantages include small storage space requirement and ease of
implementation. To compare similarity between two time series data,
Dynamic Time Warping distance measure with Sakoe-Chiba band \cite{108244}
is used.

In this paper, we extend this previous work further. Instead of using
Sakoe-Chiba band, R-K bands \cite{RatanamahatanaK04} are used and
are assigned to each user's templates with calculated threshold values.
This novel hand geometry verification system significantly increases
overall system performances, especially, reduction in False Acceptance
Rate (FAR) and False Rejection Rate (FRR) at Equal Error Rate (EER).

The rest of the paper is organized as follows. Our proposed hand geometry
verification system is described in Section 2. Section 3 discusses
our evaluation method and shows the experimental results. Finally,
in Section 4, we conclude our work and provide some suggestions.

\Section{Hand Geometry Verification System}

\begin{figure}
\noindent \begin{centering}
\begin{tabular}{c}
\includegraphics[width=8cm]{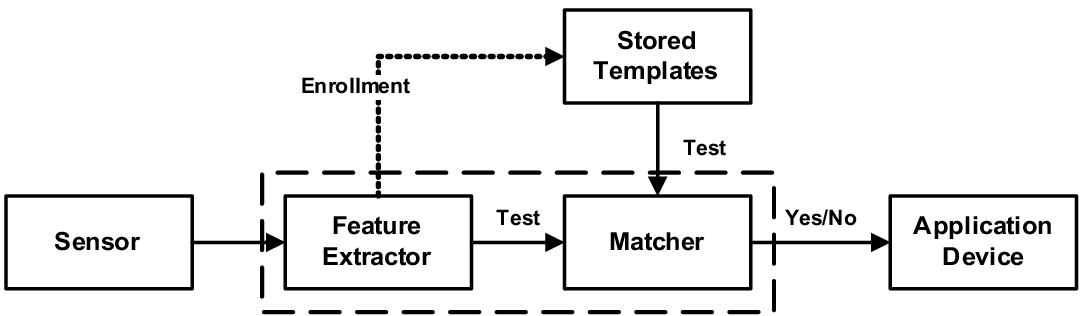}\tabularnewline
(a)\tabularnewline
\tabularnewline
\includegraphics[width=8cm]{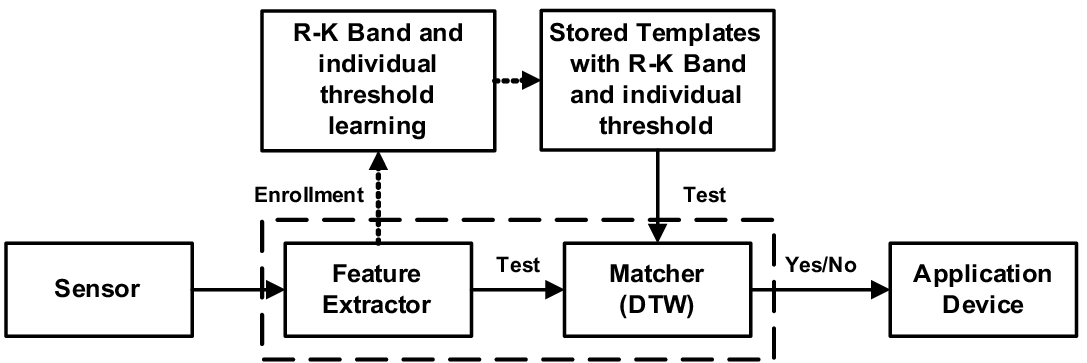}\tabularnewline
(b)\tabularnewline
\end{tabular}
\par\end{centering}

\caption{Overviews of (a) a typical biometric verification system and (b) our
proposed hand geometry verification system.}
\label{Flo:fig2}
\end{figure}

Typical verification system usually consists of five major components
\cite{JainRP06}, i.e., a sensor, a feature extractor, a matcher,
a stored template, and an application device with two functions \textendash{}
enrollment and testing, as shown in Figure \ref{Flo:fig2} (a). First,
data are acquired from a sensor, and then the input data are sent
to a feature extractor to transform the data to numerical features.
If the system is in an enrollment step, input features (from the feature
extractor) will be added to a database as a stored template labeled
with a user's identity. During the test phase, a matcher retrieves
the stored templates corresponding only to the claimed username, and
a distance measure is used to calculate similarity between the input
features and these retrieved templates. If the distance is less than
a pre-defined threshold, the system accepts; otherwise, it rejects.

Our proposed system, extended from the typical verification system,
consists of six components \textendash{} a hand scanner, a feature
extractor, R-K band and user's individual threshold learning, stored
templates with R-K band and user's individual threshold, a matcher,
and an application device. A scanner first acquires an input hand
image, and then time series data is extracted from the input image
using the feature extractor; a centroid-based \cite{KeoghWXLV06}
or an angle-based \cite{Guandi2002} technique is used for the time
series conversion. After extraction has finished and the system has
been in the enrollment step, an R-K band and user's individual threshold
is learned and stored. In the testing step, user's templates with
the learned R-K band and user's threshold are retrieved. The matcher
uses DTW distance measure with this learned R-K band to find the best
match within this stored set of templates. Finally, the measured distance
is compared to that particular user's threshold multiplied by a system-wide
threshold to grant or deny access to an application device. An overview
of our proposed system is shown in Figure \ref{Flo:fig2} (b).

\SubSection{Feature Extractor}

\begin{figure}
\noindent \begin{centering}
\includegraphics[width=6cm]{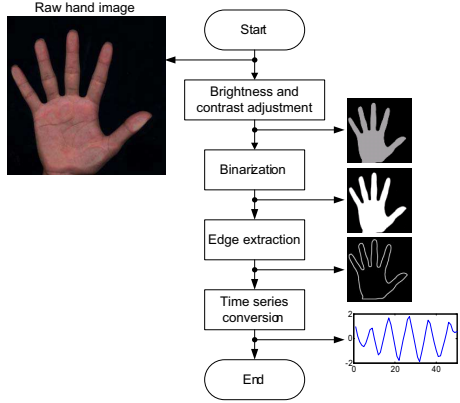}
\par\end{centering}

\caption{Steps of converting a raw hand image to time series data.}
\label{Flo:fig3}
\end{figure}

The feature extractor \cite{NiennattrakulWR07} (shown in Figure \ref{Flo:fig3})
converts acquired images to time series according to the following
steps, i.e., brightness and contrast adjustment, binarization, edge
extraction, and time series conversion.

\textbf{Brightness and Contrast Adjustment.} After an original color
raw image is obtained, it will be transformed into a grayscaled image.
Then, the system will adjust its brightness and contrast to the quality
that is suitable for the binarization step.

\textbf{Binarization.} After brightness and contrast adjustment, some
image pixels may not be pure black or white. So, we binarize each
pixel into solid black or white, represented by '0' and '1', respectively.
The binarization function is defined as follows 

\begin{equation}
B(x,y)=\left\{ \begin{array}{cc}
1 & \mathrm{if}\, I_{xy}\geq t\\
0 & \mathrm{otherwise}\end{array}\right.\label{eq:binarization}\end{equation}

\noindent where $I_{xy}$ is the intensity ranging from 0 to 1 at
pixel $(x,y)$, and $t$ is the specified threshold for binarization.
The default threshold value for the grayscale image is simply set
to 0.5.

\textbf{Edge Extraction.} The goal in this step is to find an edge
sequence from a binarized hand image by using boundary extraction
algorithm \cite{Guandi2002}. The algorithm precisely starts scanning
of each pixel from the top left of the image to the bottom right of
the image. Once it finds the first black pixel, it stops scanning,
and then traces along the edge in a clockwise direction until it returns
to the starting pixel.

\begin{figure}
\noindent \begin{centering}
\begin{tabular}{c}
\includegraphics[height=2cm]{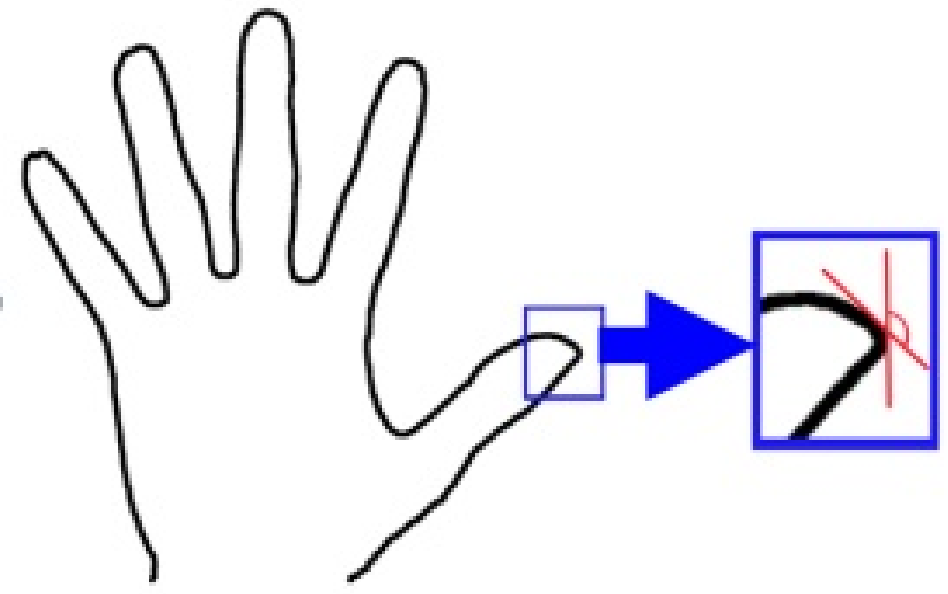}\tabularnewline
\includegraphics[width=3cm]{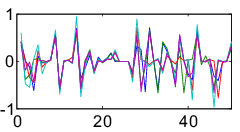}\tabularnewline
(a)\tabularnewline
\tabularnewline
\includegraphics[height=2cm]{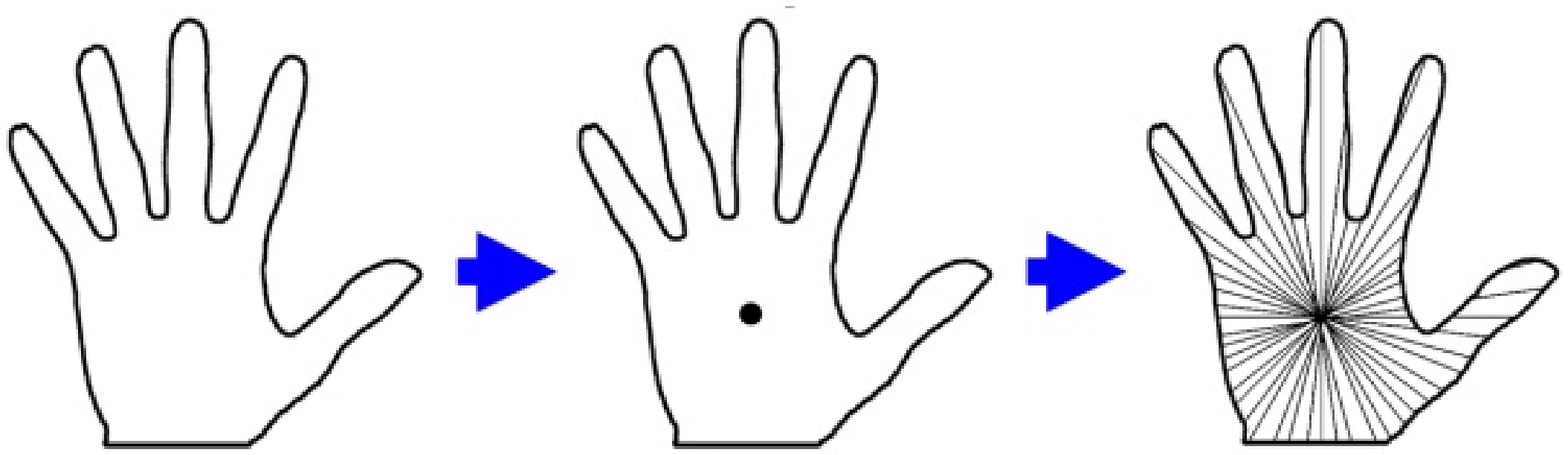}\tabularnewline
\includegraphics[width=3cm]{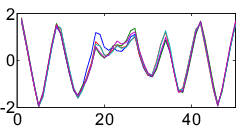}\tabularnewline
(b)\tabularnewline
\end{tabular}
\par\end{centering}

\caption{General ideas of (a) an angle-based conversion technique and (b) a
centroid-based conversion technique.}
\label{Flo:fig4}
\end{figure}

\textbf{Time Series Conversion.} In this step, we calculate the edge
sequence, and transform hand's shape into time series data by using
two techniques, i.e., an angle-based technique \cite{Guandi2002}
and a centroid-based technique \cite{KeoghWXLV06}. In the angle-based
technique, for each pixel index $i$, we create two tangent lines
\textendash{} forward and backward tangents. The forward line is created
by drawing a straight line from the pixel index $i$ to the pixel
index $i+\delta$, and the backward line is created by drawing a straight
line from the pixel index $i$ to the pixel index $i-\delta$. Note
that the $\delta$ value depends on the size of the image; the larger
the $\delta$ value, the smoother the time series, and vice versa.
In our experiment, the default value of $\delta$ is set to 10. After
that we record the angle formed by these two lines as time series'
amplitude, as shown in Figure \ref{Flo:fig4} (a). In the centroid-based
technique, we first locate the hand's centroid. Once the centroid
is obtained, we simply plot the Euclidean distance from each pixel
position to the centroid position. The general idea of centroid-based
conversion is shown in Figure \ref{Flo:fig4} (b).

\SubSection{Dynamic Time Warping (DTW) Distance Measure}

Dynamic Time Warping (DTW) \cite{BerndtC94,RatanamahatanaK05} distance
measure is a well-known similarity measure based on shape. It uses
a dynamic programming technique to find all possible warping paths,
and selects the one with the minimum distance between two time series.
To calculate the distance, it first creates a distance matrix, where
each element in the matrix is a cumulative distance of the minimum
of three surrounding neighbors. Suppose we have two time series, a
sequence $Q=\left\langle q_{1},q_{2},\ldots,q_{i},\ldots,q_{n}\right\rangle $
and a sequence $C=\left\langle c_{1},c_{2},\ldots,c_{j},\ldots,c_{m}\right\rangle $.
First, we create an $n$-by-$m$ matrix where every $(i,j)$ element
of the matrix is the cumulative distance of the distance at $(i,j)$
and the minimum of three neighboring elements, where $1\leq i\leq n$
and $1\leq j\leq m$. We can define the $(i,j)$ element, $\gamma_{ij}$,
of the matrix as: 

\begin{equation}
\gamma_{i,j}=d_{ij}+\min\left\{ \gamma_{i-1,j-1},\gamma_{i-1,j},\gamma_{i,j-1}\right\} \label{eq:dtw1}\end{equation}

\noindent where $d_{ij}=(c_{i}-q_{j})^{2}$ is the squared distance
of $q_{i}$ and $c_{j},$ and $\gamma_{i,j}$ is the summation of
$d_{ij}$ and the minimum cumulative distance of three elements surrounding
the $(i,j)$ element. Then, to find an optimal path, we choose the
path that yields a minimum cumulative distance at $(n,m)$, which
is defined as:

\begin{equation}
DTW(Q,C)=\underset{\forall W\in\mathbb{W}}{\min}\left\{ \sqrt[p]{\overset{K}{\underset{k=1}{\sum}}d_{w_{k}}}\right.\label{eq:dtw2}\end{equation}

\noindent where $\mathbb{W}$ is a set of all possible warping paths,
$w_{k}$ is $(i,j)$ at $k^{th}$ element of a warping path, and $K$
is the length of the warping path.

\begin{figure}
\noindent \begin{centering}
\begin{tabular}{cc}
\includegraphics[width=3cm]{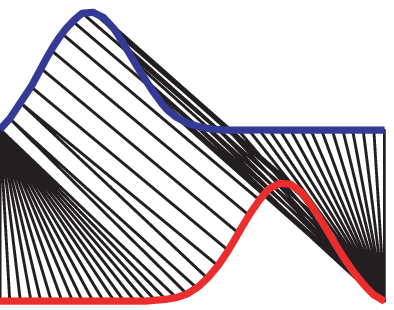} & \includegraphics[width=3cm]{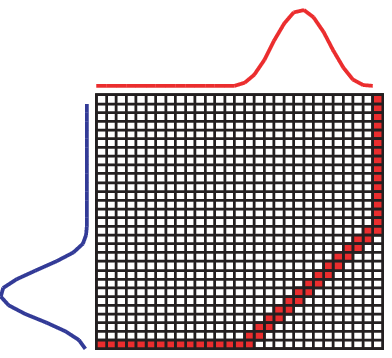}\tabularnewline
\end{tabular}
\par\end{centering}

\caption{DTW without using global constraint may introduce an unwanted warping.}
\label{Flo:fig5}
\end{figure}

\begin{figure}
\noindent \begin{centering}
\begin{tabular}{cc}
\includegraphics[width=3cm]{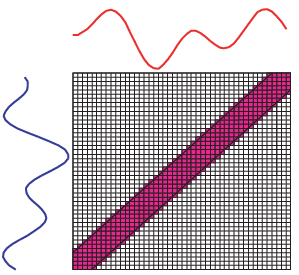} & \includegraphics[width=3cm]{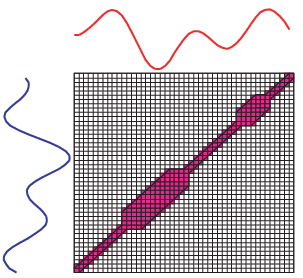}\tabularnewline
(a) & (b)\tabularnewline
\end{tabular}
\par\end{centering}

\caption{Global constraint examples of (a) Sakoe-Chiba band and (b) Ratanamahatana-Keogh
band.}
\label{Flo:fig6}
\end{figure}

In reality, DTW may not give the best mapping according to our need
because it will try its best to find the minimum distance though it
may generate an unwanted path. For example, in Figure \ref{Flo:fig5},
without a global constraint, DTW will find its optimal mapping between
two time series. However, in many cases, this is probably not what
we intend, where the two time series are expected to be of different
classes. We can resolve this problem by limiting permissible warping
paths using a global constraint. A well-known global constraints,
Sakoe-Chiba band (S-C band) \cite{108244} and a recent representation,
Ratanamahatana-Keogh band (R-K band) \cite{RatanamahatanaK04}, have
been proposed. Figure \ref{Flo:fig6} shows an example for each type
of the constraints.

\SubSection{R-K Band Learning}

Ratanamahatana-Keogh band (R-K band) \cite{RatanamahatanaK04} is
a general model of a global constraint specified by a one-dimensional
array $R$, i.e., $R=\left\langle r_{1},r_{2},\ldots,r_{i},\ldots,r_{n}\right\rangle $
where $n$ is the length of time series, and $r_{i}$ is the height
above the diagonal in $y$ direction and the width to the right of
the diagonal in $x$ direction. Each $r_{i}$ value is arbitrary,
therefore R-K band is also an arbitrary-shape global constraint, as
shown in Figure \ref{Flo:fig6} (b). Note that when $r_{i}=0$, where
$1\leq i\leq n$, this R-K band represents the Euclidean distance,
and when $r_{i}=n$, $1\leq i\leq n$, this R-K band represents the
classic DTW distance with no global constraint. The R-K band is also
able to represent the S-C band by giving all $r_{i}=c$, where $c$
is the width of a global constraint. Moreover, the R-K band is a multi
band model which can effectively be used to represent one band for
each class of the data. This flexibility is a great advantage; however,
the higher the number of classes, the larger the time complexity,
as we have to search through such a large space. 

Since determining the optimal R-K band for each training set is quite
computationally intensive, a hill climbing and heuristic functions
have been introduced to guide which part of space should be evaluated.
A space is defined as a segment of a band to be increased or decreased.
In the original work, two heuristic functions, accuracy metric and
distance metric, are used to evaluate a state. The accuracy metric
is evaluated from the training accuracy using leave-one-out 1-NN,
and the distance metric is a ratio of the mean DTW distances of correctly
classified and incorrectly classified objects.

Two searching directions are considered, i.e., a forward search, and
a backward search. In a forward search, we start from the Euclidean
distance (all $r_{i}$ values in $R$ equal to 0), and parts of the
band are gradually increased in each searching step. In the case where
two bands have the same heuristic value, a wider band is selected.
On the other hand, in a backward search, we start from a very large
band (all $r_{i}$ values in $R$ equal to $n$, where $n$ is the
length of the time series), and parts of the band are gradually decreased
in each searching step. If two bands have the same heuristic value,
the tighter band is chosen.

\begin{figure}
\noindent \begin{centering}
\includegraphics[width=6cm]{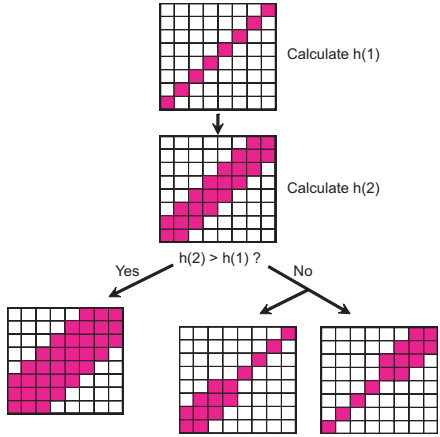}
\par\end{centering}

\caption{An illustration of the concept in R-K band forward searching algorithm.
\cite{RatanamahatanaK04} }
\label{Flo:fig7}
\end{figure}

\begin{table}
\caption{The pseudo code for multiple R-K bands learning.}

\noindent \begin{centering}
\begin{tabular}{cl}
\multicolumn{2}{l}{\noun{Function} {[}$band${]} = \noun{Learning}{[}$T$,$threshold${]}}\tabularnewline
\midrule
1 & \emph{$N$}= size of \emph{$T$};\tabularnewline
2 & \emph{$L$}= length of data in $T$;\tabularnewline
3 & initialize $band_{i}$ for $i$ = 1 to $c$;\tabularnewline
4 & foreachclass $i$ = 1 to $c$\tabularnewline
5 & ~~~enqueue(1, $L$, $Queue_{i}$);\tabularnewline
6 & endfor\tabularnewline
7 & \emph{best\_evaluate} = evaluate($T$, $band$);\tabularnewline
8 & while !empty($Queue$)\tabularnewline
9 & ~~~foreachclass $i$ = 1 to $c$\tabularnewline
10 & ~~~~~~if !empty($Queue_{i}$)\tabularnewline
11 & ~~~~~~~~~{[}$start$, $end${]} = dequeue($Queue_{i}$)\tabularnewline
12 & ~~~~~~~~~$adjustable$ = adjust($bandi$, $start$, $end$);\tabularnewline
13 & ~~~~~~~~~if $adjustable$\tabularnewline
14 & ~~~~~~~~~~~~$evaluate$= evaluate($T$, $band$);\tabularnewline
15 & ~~~~~~~~~~~~if $evaluate$ > \emph{best\_evaluate}\tabularnewline
16 & ~~~~~~~~~~~~~~~\emph{best\_evaluate} = $evaluate$;\tabularnewline
17 & ~~~~~~~~~~~~~~~enqueue($start$, $end$, $Queue_{i}$);\tabularnewline
18 & ~~~~~~~~~~~~else\tabularnewline
19 & ~~~~~~~~~~~~~~~undo\_adjustment($band_{i}$, $start$,
$end$);\tabularnewline
20 & ~~~~~~~~~~~~~~~if ($start$ \textendash{} $end$) /
2 \ensuremath{\ge} $threshold$\tabularnewline
21 & ~~~~~~~~~~~~~~~~~~enqueue($start$, $mid$-1, $Queue_{i}$);\tabularnewline
22 & ~~~~~~~~~~~~~~~~~~enqueue($mid$, $end$, $Queue_{i}$);\tabularnewline
23 & ~~~~~~~~~~~~~~~endif\tabularnewline
24 & ~~~~~~~~~~~~endif\tabularnewline
25 & ~~~~~~~~~endif\tabularnewline
26 & ~~~~~~endif\tabularnewline
27 & ~~~endfor\tabularnewline
28 & endwhile\tabularnewline
\bottomrule
\end{tabular}
\par\end{centering}

\label{Flo:table2}
\end{table}

Our learning algorithm first starts from enqueuing the starting- and
ending-parts of the R-K Band. In each iteration, these values are
dequeued and used as a boundary for a band expansion/reduction. The
adjusted band is then evaluated. If a heuristic value is higher than
the current best heuristic value, the same start and end values are
enqueued. If not, this part is further divided into two equal subparts
before being enqueued, as shown in Figure \ref{Flo:fig7}. The iterations
are continued until a termination condition is met. Table \ref{Flo:table2}
shows the pseudo code for this multiple R-K bands learning.

\Section{Experimental Evaluation}

\begin{figure}
\noindent \begin{centering}
\includegraphics[width=6cm]{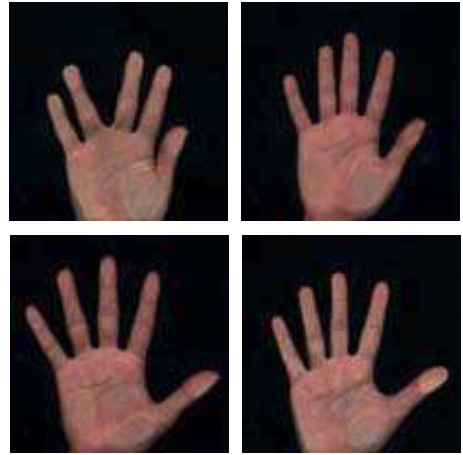}
\par\end{centering}

\caption{Examples of input hand images from four different people.}
\label{Flo:fig8}
\end{figure}

The dataset is collected from 21 different persons, 6-7 images per
person, in a total of 128 images, using a color scanner with 1200x1200
pixel resolution. Examples of input hand images are shown in Figure
\ref{Flo:fig8}.

\begin{figure}
\noindent \begin{centering}
\begin{tabular}{cc}
\includegraphics[width=4cm]{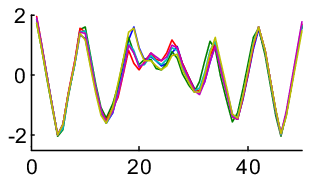} & \includegraphics[width=4cm]{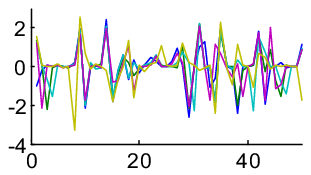}\tabularnewline
\includegraphics[width=4cm]{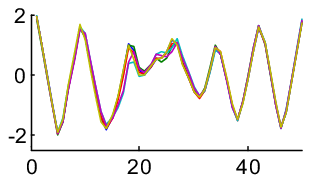} & \includegraphics[width=4cm]{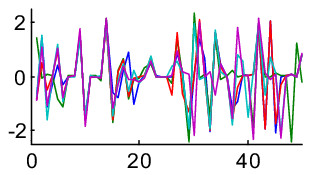}\tabularnewline
(a) & (b)\tabularnewline
\end{tabular}
\par\end{centering}

\caption{Time series data examples after the conversion using (a) the centroid-based
technique and (b) the angle-based technique.}
\label{Flo:fig9}
\end{figure}

After that all images are converted into time series data by applying
two time series conversion techniques \textendash{} a centroid-based
conversion technique and an angle-based conversion technique, and
then time series data are downsampled to 50 data points. Examples
of converted time series data using the centroid-based technique and
the angle-based technique are shown in Figure \ref{Flo:fig9} (a)
and (b), respectively.

To evaluate the system's performance, we use three well-known measurements,
i.e., False Rejection Rate (FRR), False Acceptance Rate (FAR), and
Total Success Rate (TSR). They can be calculated by the following
equations.

\begin{equation}
FRR=\frac{\#RejectGenuineClaims}{Total\#GenuineAccess}\cdot100\%\label{eq:frr}\end{equation}

\begin{equation}
FAR=\frac{\#AcceptImposterClaims}{Total\#ImposterAccess}\cdot100\%\label{eq:far}\end{equation}

\begin{equation}
TSR=\left(1-\frac{FAR+FRR}{Total\#Access}\right)\cdot100\%\label{eq:tsr}\end{equation}

\begin{figure}[h]
\noindent \begin{centering}
\begin{tabular}{c}
\includegraphics[width=6cm]{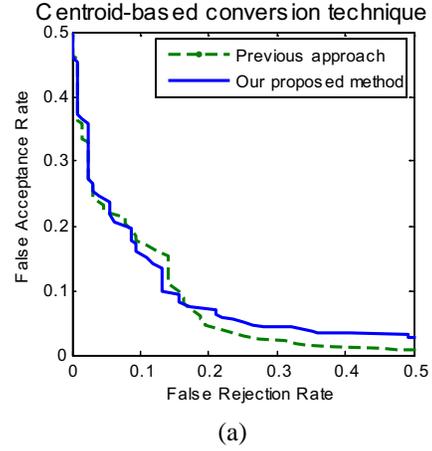}\tabularnewline
(a)\tabularnewline
\tabularnewline
\includegraphics[width=6cm]{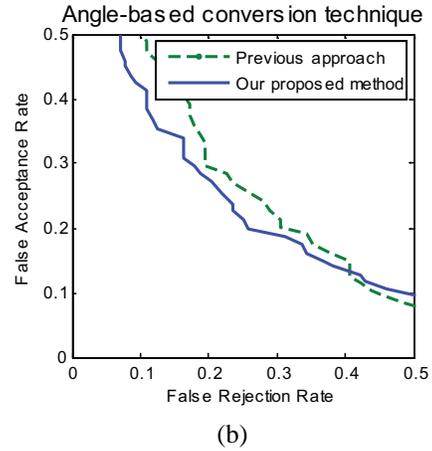}\tabularnewline
(b)\tabularnewline
\end{tabular}
\par\end{centering}

\caption{The ROC curves of our proposed system comparing with the previous
one using (a) the centroid-based conversion technique and (b) the
angle-based conversion technique.}
\label{Flo:fig10}
\end{figure}

To get the best performance, two parameters are varied \textendash{}
a systemwide (global) threshold and a conversion technique. An experimental
result at the Equal Error Rate (EER) point, where FAR equals to FRR,
is shown in Table \ref{Flo:table3}. Note that the lower the EER value,
the higher the accuracy of the biometric system. To further illustrate
our overall performance, we also show their ROC (Receive Operating
Characteristic) curves (shown in Figure \ref{Flo:fig10}). At the
EER, our proposed method gets better performance (lower FAR, lower
FRR, and higher TSR) in all evaluation metrics.

\begin{table}
\noindent \begin{centering}
\begin{tabular}{|c|c|c|}
\cline{2-3} 
\multicolumn{1}{c|}{} & \multicolumn{2}{c|}{\textbf{\footnotesize Previous approach}}\tabularnewline
\cline{2-3} 
\multicolumn{1}{c|}{} & \textbf{\footnotesize Centroid-based technique} & \textbf{\footnotesize Angle-based technique}\tabularnewline
\hline
\textbf{\footnotesize FAR} & {\footnotesize 14.06\%} & {\footnotesize 25.59\%}\tabularnewline
\hline 
\textbf{\footnotesize FRR} & {\footnotesize 14.30\%} & {\footnotesize 25.78\%}\tabularnewline
\hline 
\textbf{\footnotesize TSR} & {\footnotesize 85.71\%} & {\footnotesize 72.99\%}\tabularnewline
\hline
\multicolumn{1}{c}{} & \multicolumn{1}{c}{} & \multicolumn{1}{c}{}\tabularnewline
\cline{2-3} 
\multicolumn{1}{c|}{} & \multicolumn{2}{c|}{\textbf{\footnotesize Proposed method}}\tabularnewline
\cline{2-3} 
\multicolumn{1}{c|}{} & \textbf{\footnotesize Centroid-based techniques} & \textbf{\footnotesize Angle-based techniques}\tabularnewline
\hline 
\textbf{\footnotesize FAR} & {\footnotesize 11.72\%} & {\footnotesize 23.67\%}\tabularnewline
\hline 
\textbf{\footnotesize FRR} & {\footnotesize 11.91\%} & {\footnotesize 23.43\%}\tabularnewline
\hline 
\textbf{\footnotesize TSR} & {\footnotesize 88.10\%} & {\footnotesize 76.33\%}\tabularnewline
\hline
\end{tabular}
\par\end{centering}

\caption{The comparison of FAR, FRR, and TSR among different approaches at
EER.}
\label{Flo:table3}
\end{table}

\Section{Conclusion}

We have demonstrated our novel hand geometry verification system by
using time series data as an image representation and comparing time
series using Dynamic Time Warping distance measure with the R-K band.
In the proposed system, two time series conversion techniques are
applied, i.e., the centroid-based conversion technique and the angle-based
technique. Our experiment reveals that the centroid-based technique
generally outperforms the angle-based technique by achieving lower
EER and higher TSR. Finally, by comparing the result with the previous
approach, our proposed system gets much higher performance.

For future work, this hand geometry verification system can be extended
to be a multi-biometric system. The fingerprint and palm print can
be verified together in order to boost up the system's performance.

\bibliographystyle{latex8}

\end{document}